\documentclass[A4,12pt]{article}

\usepackage{cite}

\newcommand{\iz}[0]{{{}^1\!\!/\!{}_2 }}

\title{Cubic interaction vertices \\
for massless higher spin supermultiplets in $d=4$}

\author{M.V. Khabarov\thanks{maksim.khabarov@ihep.ru},
Yu.M. Zinoviev\thanks{Yurii.Zinoviev@ihep.ru}
\\[0.5cm]
\it{\small Institute for High Energy Physics of National
Research Center "Kurchatov Institute"} \\
\it{\small Protvino, Moscow Region, 142281, Russia}}

\date{}

\begin{document}

\maketitle

\begin{abstract}
We construct a range of supersymmetric cubic vertices for three
massless higher spin supermultiplets in the four-dimensional
space. We use frame-like multispinor formalism, which allows to avoid
most of the technical difficulties and provides a uniform description
for bosons and fermions. Our work is based on the so-called
Fradkin-Vasiliev formalism for construction of the cubic vertices,
which requires the non-zero cosmological constant. Thus we first
construct the vertices in $AdS$ space and then consider the flat
limit. We show that the AdS supersymmetric vertex is a sum of four
elementary vertices for supermultiplet components, while one of the
vertices vanishes in the flat limit in agreement with the Metsaev's
classification.
\end{abstract}

\thispagestyle{empty}
\newpage
\setcounter{page}{1}

\section{Introduction}

Recently, a general classification of cubic interaction vertices for
massless higher spin supermultiplets was developed in \cite{Met19a}
for $N=1$ supersymmetry and in \cite{Met19b} for extended $N > 1$
ones. In this work we deal with the case of $N=1$ supersymmetry,
so let us discuss the results of \cite{Met19a}.

First of all, let us briefly remind the classification of cubic
vertices for massless higher spin fields in $d=4$ \cite{Met05,Met07b}.
There exist three types of such vertices. Vertices of the type I have
$N=s_1+s_2+s_3$ derivatives and we call them trivially gauge invariant
ones because they can be written as (schematically):
$$
{\cal L}_1 \sim {\cal R}_1 {\cal R}_2 {\cal R}_3
$$
where ${\cal R}_i$ --- gauge invariant field strengths (curvatures)
having $s_i$ derivatives. In turn, the vertices of type II have
$N=s_1+s_2-s_3$ derivatives in the bosonic case or $N=s_1+s_2-s_3-1$
in the fermionic one. They can be subdivided into two sub-types (here
and in what follows we always assume that spins are ordered as $s_1
\ge s_2 \ge s_3$):
\begin{eqnarray*}
{\cal L}_1 \sim \Phi_1 {\cal R}_2 {\cal R}_3 &\qquad& s_1 \ge s_2+s_3
\\
{\cal L}_1 \sim \Phi_1 \Phi_2 {\cal R}_3 &\qquad& s_1 < s_2 + s_3
\end{eqnarray*}
We call the first sub-type abelian vertices, while the second sub-type
(whose spins satisfy the so called strict triangular inequality)
non-abelian ones. At last, the vertices of type III have 
$N=s_1-s_2-s_3$ derivatives and are known only in the light-cone
formalism.

Now let us turn to the cubic vertices for the massless $N=1$
supermultiplets. Having in our disposal three pairs of bosons and
fermions $(B_i,F_i)$, $i=1,2,3$, we can in general construct four
independent vertices (we call them elementary), namely, one bosonic
vertex and three fermionic ones:
$$
V_0(B_1,B_2,B_3), \qquad V_1(F_2,B_1,F_3), \qquad
V_2(F_1,B_2,F_3), \qquad V_3(F_1,F_2,B_3).
$$
A supersymmetric cubic vertex is a combination of such elementary ones
(with their coupling constants appropriately adjusted) which is
invariant under the supertransformations. Recall that $N=1$
supermultiplets are characterized by superspin $Y$ which for the
massless ones coincides with the lowest spin. In table 1, based on the
results of Metsaev \cite{Met19a}, we provide all possible types of
supersymmetric vertices and the combinations of the elementary ones
they consist of (note that all $s_{1,2,3}$ in the table are integer).
\begin{table}[ht!]
\centering
\begin{tabular}{|p{0.05\textwidth}|p{0.40\textwidth}|p{0.15\textwidth}|}
\hline
Type&Supermultiplets&Vertex\\
\hline
1a&$Y_1=s_1-\iz$, $Y_2=s_2-\iz$, $Y_3=s_3$&$V_0 + V_1 + V_2$\\
\hline
1b&$Y_1=s_1$, $Y_2=s_2$, $Y_3=s_3$&$V_1+V_2+V_3$\\
\hline
2a&$Y_1=s_1-\iz$, $Y_2=s_2-\iz$, $Y_3=s_3-\iz$&$V_0+V_1+V_2$\\
\hline
2b&$Y_1=s_1$, $Y_2=s_2-\iz$, $Y_3=s_3$&$V_0+V_1+V_3$\\
\hline
2c&$Y_1+s_1$, $Y_2=s_2$, $Y_3=s_3-\iz$&$V_1+V_2+V_3$\\
\hline
3a&$Y_1=s_1-\iz$, $Y_2=s_2-\iz$, $Y_3=s_3$&$V_0+V_1+V_3$\\
\hline
3b&$Y_1=s_1$, $Y_2=s_2-\iz$, $Y_3=s_3-\iz$&$V_1+V_2+V_3$\\
\hline
3c&$Y_1=s_1$, $Y_2=s_2$, $Y_3=s_3$&$V_0+V_2+V_3$\\
\hline
\end{tabular}
\caption{Different types of the massless flat supersymmetric cubic
vertices}
\label{tab:class}
\end{table}

As can be seen from the table 1, all supersymmetric vertices contain
only three elementary ones. This fact can be easily understood as
follows. In flat space the general structure of supertransformations
for the massless supermultiplets has the form (schematically):
$$
\delta B \sim F \zeta, \qquad \delta F \sim d B \zeta
$$
Then, for the supersymmetric vertex be invariant under the
supertransformations, the number of derivatives in the bosonic and
fermionic elementary vertices must satisfy
$$
N_{BBB} = N_{BFF} + 1
$$
Now let us consider three arbitrary supermultiplets 
$(s_i,s_i+\epsilon_i)$, $i=1,2,3$, where $\epsilon_i= \pm \iz$. 
For simplicity we assume that the spins are all distinct:
$s_1>s_2>s_3$.
Here we restrict ourselves with the type 2 vertices we are mostly
interested in this work. Then it is easy to calculate the number of
derivatives in each elementary vertex:
\begin{eqnarray*}
N_0 &=& s_1+s_2-s_3 \\
N_1 &=& s_1+s_2-s_3+\epsilon_2-\epsilon_3-1 \\
N_2 &=& s_1+s_2-s_3+\epsilon_1-\epsilon_3-1 \\
N_3 &=& s_1+s_2-s_3+\epsilon_1+\epsilon_2-1
\end{eqnarray*}
One can easily see that the relation $N_{BBB}=N_{BFF}+1$ cannot be
fulfilled by all four vertices simultaneously; hence, one of the
vertices has the "wrong" number of derivatives and must be absent.
There are four possible cases depending on which vertex is absent, and
in each case, the parameters $\epsilon_i$ are fixed. We consider the
cases in turn. \\
I) $V_0+V_1+V_2$: $\epsilon_{1,2}=-\iz$, $\epsilon_3=\iz$. This
corresponds to type 2a.\\
II) $V_0+V_1+V_3$: $\epsilon_{1,3}=\iz$, $\epsilon_2=-\iz$. This
corresponds to type 2b. \\
III) $V_0+V_2+V_3$: $\epsilon_1=-\iz$, $\epsilon_{2,3}=\iz$. This also
corresponds to type 2b, but with the roles of first and second
supermultiplets interchanged.\\
IV) $V_1+V_2+V_3$: $\epsilon_{1,2}=\iz$, $\epsilon_3=-\iz$. This
corresponds to type 2c.

The classification of the supersymmetric vertices has been developed in
\cite{Met19a} in the light-cone formalism. As for the Lorentz
covariant formulation, till now just a few non-trivial examples were
constructed using the superfield formalism. In two papers 
\cite{BGK17} and \cite{BGK18b} the cubic interactions for one higher
spin supermultiplets with two chiral supermultiplets $Y_2=Y_3=0$ were
constructed for half-integer and integer superspins, correspondingly.
In \cite{BGK18a} the cubic interaction for arbitrary half-integer
superspin $Y=s_1+\iz$ with two equal superspins $Y_2=Y_3$ (which may
be integer or half-integer) were constructed. The two lower superspin
multiplets enter through their gauge invariant field strengths, so
this
gives examples of the type 2ab abelian vertices. The type 2c vertices
are absent just because the authors considered only the case where
lower superspins are equal, but, in-principle, they also can be
constructed with the same technique. At last, in \cite{GK19} the
authors considered the case of arbitrary integer superspin $Y_1=s_1$
and again two equal superspins $Y_2=Y_3$ (which also may be integer or
half-integer). From the number of derivatives it follows that these
vertices belong to the class of trivially gauge invariant ones 1ab.

In this work we provide an explicit construction for the
supersymmetric cubic vertices of type 2abc when all spins satisfy the
strict triangular inequality (so our results are complementary to that
of \cite{BGK18a}). The construction is heavily based on our previous
work \cite{KhZ20a} devoted to the construction of (what we now call)
elementary bosonic and fermionic vertices using a so-called
Fradkin-Vasiliev formalism \cite{FV87,FV87a,Vas11}. Let us briefly
remind the
procedure.

Recall, that in a frame-like formalism each massless bosonic or
fermionic higher spin particle is described by a set of gauge fields
(one-forms) $\Phi$ (physical, auxiliary and extra ones). For each
field, a gauge invariant curvature (two-form) ${\cal R}$ can be
constructed. At last, in $AdS$ space the free Lagrangian can be
rewritten in the explicitly gauge invariant form as
$$
{\cal L}_0 \sim \sum_k a_k {\cal R}_k {\cal R}_k
$$
The first step in constructing the cubic vertex in the
Fradkin-Vasiliev formalism is to find quadratic deformations for all
curvatures $\Delta {\cal R} \sim \Phi\Phi$ such that the deformed
curvatures $\hat{\cal R} = {\cal R} + \Delta{\cal R}$ transform
covariantly under the gauge transformations $\delta \hat{\cal R} \sim
{\cal R} \eta$. On the second step, one takes the sum of the three
Lagrangians and replace the initial curvatures with the deformed ones,
requiring that the result be gauge invariant (on-shell). The cubic
terms of the deformed Lagrangian constitutes the desired (on-shell)
gauge invariant cubic vertex. As described, such formalism works only
in $AdS$ space with non-zero cosmological constant, but, as we have
shown in \cite{KhZ20a}, the resulting cubic vertices admits (on-shell)
a non-singular flat limit, providing us with the flat space cubic
vertices as well.

Our current work is a straightforward generalization of this procedure
to the case of supersymmetric cubic vertices. The main idea is really
very simple. We consider quadratic deformations for all bosonic and
fermionic curvatures corresponding to all four elementary vertices
simultaneously. Besides the usual requirement that the deformed
curvatures must transform covariantly under the gauge transformations,
we also require that under the supertransformations they transform
exactly in the same way as the initial ones. This ensures that the
deformed Lagrangian and hence a cubic vertex will be (on-shell)
invariant under the supertransformations. Here we also begin with the
$AdS$ space with non-zero cosmological constant and then consider the
flat limit. We will see, that in $AdS$ all supersymmetric vertices
necessarily contain all four elementary ones, while one of them 
always vanishes in the flat limit in complete agreement with Metsaev's
results.

The paper is organized as follows. In Section 2, we provide the
necessary background information about the frame-like description of
the massless higher spin fields and supermultiplets. In Section 3, we
construct the AdS vertices, while in Section 4 we provide examples of
the flat space supersymmetric vertices. Notation and conventions as
well as some technical details are collected in a pair of appendices.

\section{Free massless HS particles and supermultiplets}

In this section we provide all necessary information on the massless
higher spin bosons, fermions and supermultiplets in the frame-like
multispinor formalism.

\subsection{Boson}

To describe the massless spin-$s \ge 2$ boson in the frame-like
formalism, one needs the physical field
$H^{\alpha(s-1)\dot\alpha(s-1)}$, a pair of auxiliary fields
$\Omega^{\alpha(s)\dot\alpha(s-2)}$,
$\Omega^{\alpha(s-2)\dot\alpha(s)}$ and (for $s \ge 3$) a set of 
so-called extra fields $\Omega^{\alpha(s-1+m)\dot\alpha(s-1-m)}$, 
$2 \le |m| \le s-1$. All fields are gauge one-forms. If we iterate
over the complete set of fields, we denote them as
$\Omega^{\alpha(s-1+m)\dot\alpha(s-1-m)}$, $|m| \le s-1$ for brevity,
assuming that for $m=0$ $\Omega^{\alpha(s-1+m)\dot\alpha(s-1-m)}
\equiv H^{\alpha(s-1)\dot\alpha(s-1)}$.

The gauge transformations for the fields have the form:
\begin{eqnarray}
\delta  H^{\alpha(s-1)\dot\alpha(s-1)} &=& D 
\eta^{\alpha(s-1)\dot\alpha(s-1)}
+ e^\alpha{}_{\dot\beta} 
\eta^{\alpha(s-2)\dot\beta\dot\alpha(s-1)}
+ e_\beta{}^{\dot\alpha} 
\eta^{\alpha(s-2)\beta\dot\alpha(s-1)},
 \nonumber \\ 
\delta  \Omega^{\alpha(s-1+m)\dot\alpha(s-1-m)} &=& D 
\eta^{\alpha(s-1+m)\dot\alpha(s-1-m)}
+ \lambda^2 e^\alpha{}_{\dot\beta} 
\eta^{\alpha(s-2+m)\dot\beta\dot\alpha(s-1-m)}
 \nonumber \\ 
&& + e_\beta{}^{\dot\alpha} 
\eta^{\alpha(s-1+m)\beta\dot\alpha(s-2-m)}, \qquad 0<m<s-1,
 \\ 
\delta  \Omega^{\alpha(2s-2)} &=& D  \eta^{\alpha(2s-2)}
+ \lambda^2 e^\alpha{}_{\dot\alpha} 
\eta^{\alpha(2s-3)\dot\alpha}. \nonumber
\end{eqnarray}
For each field $\Omega^{\alpha(s-1+m)\dot\alpha(s-1-m)}$, a gauge
invariant two-form can be constructed:
\begin{eqnarray}
 {\mathcal{T}}^{\alpha(s-1)\dot\alpha(s-1)} &=& D 
H^{\alpha(s-1)\dot\alpha(s-1)}
+ e^\alpha{}_{\dot\beta} 
\Omega^{\alpha(s-2)\dot\beta\dot\alpha(s-1)}
+ e_\beta{}^{\dot\alpha} 
\Omega^{\alpha(s-1)\beta\dot\alpha(s-2)},
 \nonumber \\ 
 {\mathcal{R}}^{\alpha(s-1+m)\dot\alpha(s-1-m)} &=& D 
\Omega^{\alpha(s-1+m)\dot\alpha(s-1-m)}
+ \lambda^2 e^\alpha{}_{\dot\beta} 
\Omega^{\alpha(s-2+m)\dot\beta\dot\alpha(s-1-m)}
 \nonumber \\ 
&& + e_\beta{}^{\dot\alpha} 
\Omega^{\alpha(s-1+m)\beta\dot\alpha(s-2-m)}, \qquad 0<m<s-1,
  \\ 
 {\mathcal{R}}^{\alpha(2s-2)} &=& D  \Omega^{\alpha(2s-2)}
+ \lambda^2 e^\alpha{}_{\dot\alpha} 
\Omega^{\alpha(2s-3)\dot\alpha}. \nonumber
\end{eqnarray}
We refer to these two-forms as curvatures. Similarly, we assume that\\
$\mathcal{R}^{\alpha(s-1)\dot{\alpha}(s-1)} \equiv
\mathcal{T}^{\alpha(s-1)\dot{\alpha}(s-1)}$ whenever the expression 
$\mathcal{R}^{\alpha(s-1+m)\dot{\alpha}(s-1-m)}$, $|m|\le s-1$ is
encountered. It is straightforward to check that these curvatures
satisfy the following differential identities:
\begin{eqnarray}
D {\mathcal{T}}^{\alpha(s-1)\dot\alpha(s-1)} &=&
- e^\alpha{}_{\dot\beta} 
{\mathcal{R}}^{\alpha(s-2)\dot\beta\dot\alpha(s-1)}
- e_\beta{}^{\dot\alpha} 
{\mathcal{R}}^{\alpha(s-1)\beta\dot\alpha(s-2)},
 \nonumber \\ 
D  {\mathcal{R}}^{\alpha(s-1+m)\dot\alpha(s-1-m)} &=& 
- \lambda^2 e^\alpha{}_{\dot\beta} 
{\mathcal{R}}^{\alpha(s-2+m)\dot\beta\dot\alpha(s-1-m)}
 \nonumber \\ &&
- e_\beta{}^{\dot\alpha} 
{\mathcal{R}}^{\alpha(s-1+m)\beta\dot\alpha(s-2-m)}, \qquad 0<m<s-1,
 \\ 
D {\mathcal{R}}^{\alpha(2s-2)} &=&
-\lambda^2 e^\alpha{}_{\dot\alpha} 
{\mathcal{R}}^{\alpha(2s-3)\dot\alpha}. \nonumber
\end{eqnarray}

On-shell all the curvatures, except for the highest ones, i.e. $
{\mathcal{R}}^{\alpha(2s-2)}$, are zero, while the highest curvature
can be parameterized by a gauge invariant zero-form $W^{\alpha(2s)}$:
\begin{equation}
\label{riemann_weyl}
{\cal R}^{\alpha(2s-2)} = - E_{\beta(2)} W^{\alpha(2s-2)\beta(2)}
\end{equation}
In case of gravitation, $s=2$, the equation (\ref{riemann_weyl})
expresses a well-known fact that in absence of the matter, the Riemann
tensor is traceless, i.e. is equal to the Weyl tensor. We thus refer
to $W^{\alpha(2s)}$ as generalised Weyl tensor.

Let us rewrite the on-shell conditions in terms of the fields:
\begin{eqnarray*}
D \Omega^{\alpha(s-1)\dot\alpha(s-1)} &=&  - e_\beta{}^{\dot\alpha}
\Omega^{\alpha(s-1)\beta\dot\alpha(s-2)} - h.c. \\
D \Omega^{\alpha(s-1+m)\dot\alpha(s-1-m)} &=& - e_\beta{}^{\dot\alpha}
\Omega^{\alpha(s-1+m)\beta\dot\alpha(s-2-m)} + O(\lambda^2) \\
D \Omega^{\alpha(2s-2)} &=& - E_{\beta(2)} W^{\alpha(2s-2)\beta(2)}
+ O(\lambda^2) 
\end{eqnarray*}
Hence, on-shell the auxiliary field expresses the non-zero derivatives
of the physical field, the extra field 
$\Omega^{\alpha(s+1)\dot\alpha(s-3)},h.c.$ 
expresses the non-zero derivatives of the auxiliary field etc. The
field $\Omega^{\alpha(s-1+m)\dot\alpha(s-1-m)},h.c.$ thus expresses
the $m$-th derivatives of the physical field which do not vanish 
on-shell. Finally, the Weyl tensor expresses the $s$-th nonvanishing
derivatives. Whenever we talk about the number of derivatives, we
imply the number of derivatives of the physical field and count the
$m$-th field as an $m$-th derivative.

The free Lagrangian can be written in the explicitly gauge
invariant form
\begin{equation}
{\cal L}_0 = i \sum_{m=1}^{s-1} 
\frac{(-1)^s(2s-2)!}{(s-1+m)!(s-1-m)!\lambda^{2m}} 
{\cal R}_{\alpha(s-1+m)\dot\alpha(s-1-m)} 
{\cal R}^{\alpha(s-1+m)\dot\alpha(s-1-m)} + h.c.
\end{equation}
This expression exists in the non-flat space only.
Note that the torsion ${\cal T}^{\alpha(s-1)\dot\alpha(s-1)}$ is
absent in this expression. The free Lagrangian terms have at most two 
derivatives of the physical field, and thus cannot contain the extra
fields.
This constitutes the so-called extra field decoupling condition. 
Together with the normalization condition, it uniquely determines the 
coefficients in the explicitly gauge invariant form of the
Lagrangian.

\subsection{Fermion}

A massless half-integer spin-($s+\iz$), $s \ge 1$,  fermion is
described by a set of multispinor gauge one-forms
$\Psi^{\alpha(s+m)\dot\alpha(s-1-m)}+h.c.$,
$0 \le m \le s-1$, where $m=0$ correspond
to the physical field, all others being the extra ones. The general
form of the gauge transformations is similar to the bosonic case. The
only differences are the coefficients in the physical field
transformation law:
\begin{eqnarray}
\delta \Psi^{\alpha(s)\dot\alpha(s-1)} &=& D
\xi^{\alpha(s)\dot\alpha(s-1)} + e_\beta{}^{\dot\alpha}
\xi^{\alpha(s)\beta\dot\alpha(s-2)} + \lambda
e^\alpha{}_{\dot\beta} \xi^{\alpha(s-1)\dot\alpha(s-1)\dot\beta},
\nonumber \\
\delta \Psi^{\alpha(s+m)\dot\alpha(s-1-m)} &=& D
\xi^{\alpha(s+m)\dot\alpha(s-1-m)} + e_\beta{}^{\dot\alpha}
\xi^{\alpha(s+m)\beta\dot\alpha(s-2-m)} \nonumber \\
 && + \lambda^2 e^\alpha{}_{\dot\beta}
\xi^{\alpha(s-1+m)\dot\alpha(s-1-m)\dot\beta}, \qquad 
0 < m < s-1, 
\\
\delta \Psi^{\alpha(2s-1)} &=& D \xi^{\alpha(2s-1)} + \lambda^2
e^\alpha{}_{\dot\alpha} \xi^{\alpha(2s-2)\dot\alpha}. \nonumber
\end{eqnarray}

Similarly, a set of the gauge invariant two-forms, curvatures, can be
constructed:
\begin{eqnarray}
{\cal F}^{\alpha(s)\dot\alpha(s-1)} &=& D 
\Psi^{\alpha(s)\dot\alpha(s-1)} + e_\beta{}^{\dot\alpha}
\Psi^{\alpha(s)\beta\dot\alpha(s-2)} + \lambda 
e^\alpha{}_{\dot\beta} \Phi^{\alpha(s-1)\dot\alpha(s-1)\dot\beta},
\nonumber \\
{\cal F}^{\alpha(s+m)\dot\alpha(s-1-m)} &=& D
\Psi^{\alpha(s+m)\dot\alpha(s-1-m)} + e_\beta{}^{\dot\alpha}
\Psi^{\alpha(s+m)\beta\dot\alpha(s-2-m)} 
\nonumber \\
 && + \lambda^2 e^\alpha{}_{\dot\beta} 
\Psi^{\alpha(s-1+m)\dot\alpha(s-1-m)\dot\beta}, \qquad 0 < m < s-1,
\\
{\cal F}^{\alpha(2s-1)} &=& D \Psi^{\alpha(2s-1)} + \lambda^2
e^\alpha{}_{\dot\alpha} \Psi^{\alpha(2s-1)\dot\alpha}.
\nonumber 
\end{eqnarray}
Again, on-shell all these curvatures, except for the highest ones,
i.e. ${\cal F}^{\alpha(2s-1)}$, are zero, while the highest ones can
be parameterized by a gauge-invariant zero-form $ Y^{\alpha(2s+1)}$:
\begin{equation}
{\cal F}^{\alpha(2s-1)} = - E_{\alpha(2)} Y^{\alpha(2s+1)}
\end{equation}
Again, the zero-curvature conditions imply that the field
$\Phi^{\alpha(s+m)\dot\alpha(s-1-m)}$ expresses 
the $m$-th derivatives of the physical field 
$\Phi^{\alpha(s)\dot\alpha(s-1)}$ which do not vanish on-shell.

At last, the free Lagrangian can be written as
\begin{equation}
{\cal L}_0 = \sum_{m=0}^{s-1}
\frac{(-1)^{s-1}(2s-1)!}{(s+m)!(s-1-m)!\lambda^{2m}}
{\cal F}_{\alpha(s+m)\dot\alpha(s-1-m)}
{\cal F}^{\alpha(s+m)\dot\alpha(s-1-m)} + h.c.
\end{equation}
Again, this expression exists in non-flat space only.

\subsection{Supermultiplets}

A massless supermultiplet consists of a massless boson and a massless
fermion whose spins differ by $\iz$. We call the lower one of the two
spins the superspin. Then, there exist two different types of the
massless supermultiplets, namely the integer superspin and the
half-integer superspin massless supermultiplets. We first discuss the
general properties of generic free massless supermultiplets, and then
consider integer and half-integer superspin cases separately.

The most important property is the fact that the gauge invariant
curvatures for the bosonic and fermionic components of the
supermultiplets transform covariantly under the supertransformations,
showing that the gauge transformations and supertransformations agree.
Moreover, the on-shell conditions (a union of the on-shell equations
for the bosonic and fermionic components)  are also consistent with
the supertransformations.

We consider now the integer superspin-$s$. Its description requires
bosonic gauge one-forms set
$\Omega^{\alpha(s-1+m)\dot\alpha(s-1-m)}$, $|m|\le s-1$ and the
fermionic ones $\Psi^{\alpha(s+m)\dot\alpha(s-1-m)}$, $-s \le m \le
s-1$. The supertransformations have the form \cite{KhZ20}:
\begin{eqnarray}
\label{strans_23_int}
\delta H^{\alpha(s-1)\dot\alpha(s-1)} &=&
iC  \Psi^{\alpha(s-1)\dot\beta\dot\alpha(s-1)} \zeta_{\dot\beta}
+ iC \Psi^{\alpha(s-1)\beta\dot\alpha(s-1)} \zeta_\beta,
 \nonumber \\ 
\delta \Omega^{\alpha(s-1+m)\dot\alpha(s-1-m)} &=&
iC  \Psi^{\alpha(s-1+m)\beta\dot\alpha(s-1-m)} \zeta_{\beta}
 \nonumber \\ 
&& + i C \lambda  \Psi^{\alpha(s-1+m)\dot\beta\dot\alpha(s-1-m)}
\zeta_{\dot\beta}, \qquad 1 \le m \le s-1,
 \nonumber \\ 
\delta \Psi^{\alpha(s+m)\dot\alpha(s-1-m)}&=&
\tilde{C} \lambda \Omega^{\alpha(s-1+m)\dot\alpha(s-1-m)}
\zeta^{\alpha}
  \\ 
&& + \tilde{C}  \Omega^{\alpha(s+m)\dot\alpha(s-2-m)}
\zeta^{\dot\alpha}, \qquad 0 \le m \le s-2,
 \nonumber \\ 
\delta \Psi^{\alpha(2s-1)} &=& \lambda \tilde{C} 
\Omega^{\alpha(2s-2)} \zeta^{\alpha} + \tilde{C} e_{\beta\dot\alpha}
W^{\alpha(2s-1)\beta} \zeta^{\dot\alpha}. \nonumber
\end{eqnarray}
The generalized Weyl tensors $ W^{\alpha(2s)}$, $Y^{\alpha(2s+1)}$
have their own transformation laws; however, we do not need them in
the present work. Similarly, for the gauge invariant curvatures we
obtain (note additional terms proportional to the Weyl tensor 
$ W^{\alpha(2s)}$ for the fermionic curvatures 
$\mathcal{F}^{\alpha(2s-1)}$, 
$\mathcal{F}^{\alpha(2s-2)\dot{\alpha}}$, which will be important in
what follows):
\begin{eqnarray}
\delta {{\mathcal{T}}}^{\alpha(s-1)\dot\alpha(s-1)} &=&
iC  {{\mathcal{F}}}^{\alpha(s-1)\beta\dot\alpha(s-1)} \zeta_{\beta}
+ iC {\mathcal F}^{\alpha(s-1)\dot\alpha(s-1)\dot\beta}
\zeta_{\dot\beta},
 \nonumber \\ 
\delta {{\mathcal{R}}}^{\alpha(s-1+m)\dot\alpha(s-1-m)} &=&
iC  {{\mathcal{F}}}^{\alpha(s-1+m)\beta\dot\alpha(s-1-m)}
\zeta_{\beta}
 \nonumber \\ 
&& + iC\lambda 
{{\mathcal{F}}}^{\alpha(s-1+m)\dot\beta\dot\alpha(s-1-m)}
\zeta_{\dot\beta}, \qquad 1 \le m \le s-1,
 \nonumber \\ 
\delta {{\mathcal{F}}}^{\alpha(s+m)\dot\alpha(s-1-m)} &=&
\lambda\tilde{C} {{\mathcal{R}}}^{\alpha(s-1+m)\dot\alpha(s-1-m)}
\zeta^{\alpha}
  \\ 
&& + \tilde{C}  {{\mathcal{R}}}^{\alpha(s+m)\dot\alpha(s-2-m)}
\zeta^{\dot\alpha}, \qquad 0 \le m \le s-3,
 \nonumber \\ 
\delta {{\mathcal{F}}}^{\alpha(2s-2)\dot\alpha} &=&
\lambda\tilde{C} {{\mathcal{R}}}^{\alpha(2s-3)\dot\alpha}
\zeta^{\alpha}
+ \tilde{C} \big[ {{\mathcal{R}}}^{\alpha(2s-2)} + E_{\alpha(2)}
W^{\alpha(2s)}\big] \zeta^{\dot\alpha}
 \nonumber \\ 
\delta {{\mathcal{F}}}^{\alpha(2s-1)} &=& \lambda\tilde{C} 
{{\mathcal{R}}}^{\alpha(2s-2)} \zeta^{\alpha} + \tilde{C}
e_{\beta\dot\alpha}D W^{\alpha(2s-1)\beta} \zeta^{\dot\alpha}
\nonumber
\end{eqnarray}
The requirement that the sum of the bosonic and fermionic Lagrangians
be invariant under the supertransformations fixes the ratio 
$C/\tilde{C}$; under our choice of the normalization 
\begin{equation}
C = (2s-1) \tilde{C}
\end{equation}
The sign of the constants $C$, $\tilde{C}$ can be chosen arbitrarily;
we choose $C,\tilde{C}>0$.

In case of half-integer superspin-($s-\iz$), the supertransformations
have the form:
\begin{eqnarray}
\label{strans_23_halfint}
\delta H^{\alpha(s-1)\dot\alpha(s-1)} &=&
iC \Phi^{\alpha(s-1)\dot\alpha(s-2)}\zeta^{\dot\alpha} + 
iC \Phi^{\alpha(s-2)\dot\alpha(s-1)} \zeta^\alpha,
 \nonumber \\ 
\delta\Omega^{\alpha(s-1+m)\dot\alpha(s-1-m)} &=&
iC \Phi^{\alpha(s-1+m)\dot\alpha(s-2-m)} \zeta^{\dot\alpha}
 \nonumber \\ 
&& + iC \lambda
\Phi^{\alpha(s-2+m)\dot\alpha(s-1-m)} \zeta^\alpha,
\qquad 1 \le m \le s-2,
 \nonumber \\ 
\delta\Omega^{\alpha(2s-2)} &=& iC\lambda \Phi^{\alpha(2s-3)}
\zeta^\alpha + iC e_{\beta\dot\alpha} Y^{\alpha(2s-2)\beta}
\zeta^{\dot\alpha},
  \\ 
\delta\Phi^{\alpha(s+m-1)\dot\alpha(s-2-m)} &=&
\tilde{C} \lambda
\Omega^{\alpha(s-1+m)\dot\alpha(s-2-m)\dot{\beta}} \zeta_{\dot\beta}
 \nonumber \\ 
&& + \tilde{C}
\Omega^{\alpha(s-1+m)\beta\dot\alpha(s-2-m)} \zeta_{\beta}, \qquad
0 \le m \le s-1. \nonumber
\end{eqnarray}
Again, we do not need the transformation laws of generalized Weyl
tensors. For the gauge invariant curvatures we obtain (note again
special cases for bosonic curvatures $R^{\alpha(2s-2)}$,
$R^{\alpha(2s-3)\dot\alpha}$:
\begin{eqnarray}
\delta{\mathcal{T}}^{\alpha(s-1)\dot\alpha(s-1) } &=&
iC {\mathcal{F}}^{\alpha(s-1)\dot\alpha(s-2)} \zeta^{\dot\alpha}
+ iC {\mathcal F}^{\alpha(s-2)\dot\alpha(s-1)} \zeta^\alpha,
 \nonumber \\ 
\delta{\mathcal{R}}^{\alpha(s-1+m)\dot\alpha(s-1-m)} &=&
iC {\mathcal{F}}^{\alpha(s+m-1)\dot\alpha(s-2-m)}
\zeta^{\dot\alpha}
 \nonumber \\ 
&& + iC \lambda {\mathcal{F}}^{\alpha(s-2+m)\dot\alpha(s-m-1)}
\zeta^\alpha, \qquad 1 \le m \le s-3,
 \nonumber \\ 
\delta{\mathcal{R}}^{\alpha(2s-3)\dot\alpha} &=&
iC \big[{\mathcal{F}}^{\alpha(2s-3)} + E_{\beta(2)} 
Y^{\alpha(2s-3)\beta(2)} \big] \zeta^{\dot\alpha},
 \\ 
\delta{\mathcal{R}}^{\alpha(2s-2)} &=&
iC \lambda {\mathcal{F}}^{\alpha(2s-3)} \zeta^\alpha + iC
e_{\beta\dot\alpha} DY^{\alpha(2s-2)\beta} \zeta^{\dot\alpha},
 \nonumber \\ 
\delta{\mathcal{F}}^{\alpha(s-1+m)\dot\alpha(s-2-m)} &=&
\tilde{C} \lambda 
{\mathcal{R}}^{\alpha(s-1+m)\dot\alpha(s-2-m)\dot\beta}\zeta_{\dot\beta}
 \nonumber \\ 
&& + \tilde{C}
{\mathcal{R}}^{\alpha(s-1+m)\beta\dot\alpha(s-2-m)}
\zeta_{\beta}, \qquad 0 \le m \le s-1. \nonumber
\end{eqnarray}
The two factors $C$,$\tilde{C}$ are real; their ratio is fixed by
requiring the sum of the bosonic and fermionic Lagrangians be
invariant
under the supertransformations:
\begin{equation}
(2s-2) C = \tilde{C}
\end{equation}
Again, we fix their signs as $C,\tilde{C}>0$.

The algebra of supertransformations is closed, i.e. the anticommutator
of the two supertransformations with parameters $\zeta_1,\zeta_2$
gives (on-shell) a combination of Lorentz transformations and
translations:
\begin{eqnarray}
\label{anticommutator}
[\delta_1,\delta_2]\Omega^{\alpha(s-1+m)\dot\alpha(s-1-m)} &=&
iC\tilde{C} 
\bigg[\lambda \Omega^{\alpha(s-2+m)\beta\dot\alpha(s-1-m)}
\eta_\beta{}^\alpha
+ \lambda \Omega^{\alpha(s-1+m)\dot\alpha(s-2-m)\dot\beta}
\eta_{\dot\beta}{}^{\dot\alpha}
 \nonumber \\ 
&& + \lambda^2 \Omega^{\alpha(s-1+m)\beta\dot\alpha(s-2-m)}
\xi_\beta{}^{\dot\alpha}
+ \Omega^{\alpha(s-2+m)\dot\alpha(s-1-m)\dot\beta}
\xi^\alpha{}_{\dot\beta}\bigg]
\end{eqnarray}
where
$$
\eta^{\alpha(2)} = 2\zeta_2^\alpha \zeta_1^\alpha, \qquad
\eta^{\dot\alpha(2)} = 2 \zeta_1^{\dot\alpha} \zeta_2^{\dot\alpha},
\qquad \xi^{\alpha\dot\alpha} = \zeta_2^\alpha \zeta_1^{\dot\alpha} -
\zeta_2^\alpha \zeta_1^{\dot\alpha}
$$
The expressions are the same for both supermultiplets; one can obtain
similar ones for the fermions. In what follows we set
$C\tilde{C} = 1$.

\section{Supersymmetric vertices in $AdS_4$}

In this section we consider cubic interactions for massless higher
spin supermultiplets. Let us take three such supermultiplets
$(\Omega_i,\Phi_i)$, $i=1,2,3$. In what follows we assume that they
are ordered by their superspins $Y_1 \ge Y_2 \ge Y_3$. With these
fields four cubic vertices which we call elementary can be
constructed,
namely, one bosonic vertex and three fermionic ones:
$$
V_0(\Omega_1, \Omega_2, \Omega_3), \qquad
V_1(\Omega_1, \Phi_2, \Phi_3), \qquad
V_2(\Phi_1, \Omega_2, \Phi_3), \qquad
V_3(\Phi_1, \Phi_2, \Omega_3).
$$
A supersymmetric cubic vertex is a combination of the elementary
ones (including appropriate relations on their coupling constants)
invariant under the global supertransformations.

In \cite{KhZ20a}, elementary cubic vertices for massless higher spin
particles with spins satisfying the strict triangle inequality has
been constructed.  Let us briefly recall the main steps of such
construction. For concreteness, we consider the bosonic vertex; the
fermionic case is similar. We enumerate the particles with index
$i=1,2,3$ and assume that their spins $s_i$ are ordered from highest
to lowest: $s_1\ge s_2\ge s_3$. As we have seen in the previous
section, each particle is described by the set of one-forms
$\Omega_i^{\alpha(s_i-1+m)\dot\alpha(s_i-1-m)}$ and the corresponding
gauge invariant two-forms 
${\mathcal R}_i^{\alpha(s_i-1+m)\dot\alpha(s_i-1-m)}$. 

The first step is to deform curvatures by adding quadratic terms.
Without loss of generality, we consider the first particle. Its
curvatures receive quadratic corrections $\Delta \mathcal{R}_1 \sim
\Omega_2\Omega_3$ (see explicit expressions in Appendix B). The
corrections to the gauge transformations $\Delta \delta \Omega_1 \sim
\eta_2\Omega_3-\Omega_2\eta_3$ can be read immediately from the
corrections to the curvatures. The main requirement at this step is
that the deformed curvatures $\hat{\mathcal R}_1 = {\mathcal R} +
\Delta {\mathcal R}$ must transform covariantly under the gauge
transformations: 
$\delta \hat{\mathcal{R}}_1\sim \eta_2 \mathcal{R}_3 + \eta_3
\mathcal{R}_2$.
The curvature deformation procedure is carried out for each of the
three particles independently; each deformation is fixed up to a total
factor, which we denote $a$, $b$ and $c$ for the first, the second and
the third particle respectively.

The second step is to construct the interacting Lagrangian
$\mathcal{L}$. It is built from the sum of three particles
Lagrangians $\mathcal{L}_i$ expressed in curvatures, with linearized
curvatures replaced by the deformed ones. The requirement that the
Lagrangian be gauge invariant on-shell links the coefficients $a,b,c$
up to a single total factor $g$ (see Appendix B).

We turn on now to the main objective of our paper - the cubic
interaction of three massless supermultiplets. Again, the first step
is to build quadratic deformations of the curvatures. Without loss of
generality, we consider the first supermultiplet. The curvatures of
the first supermultiplets receive the deformations of the following
form:
\begin{eqnarray}
\Delta \mathcal{R}_1 &=& a_0\Delta \mathcal{R}_1(\Omega_2,\Omega_3) 
+ a_1 \Delta \mathcal{R}_1(\Phi_2,\Phi_3), 
 \nonumber \\ 
\Delta \mathcal{F}_1 &=& a_2 \Delta \mathcal{F}_1(\Omega_2,\Phi_3) +
a_3 \Delta \mathcal{F}_1(\Phi_2,\Omega_3).
\end{eqnarray}
The total factors $a_i$ of each elementary deformation are written out
explicitly here. The structure of the deformations of the second and
third supermultiplet curvatures is similar, with the coefficients
$b_i$ and $c_i$, $i=0,1,2,3$. Our main requirement here is that the
deformed curvatures transform under the supertransformations exactly
as the non-deformed ones:
$$
\delta \hat{\mathcal{R}} = \hat{\mathcal{F}} \zeta, \qquad
\delta \hat{\mathcal{F}} = \hat{\mathcal{R}} \zeta.
$$
This guarantees that the interacting Lagrangian (again constructed by
the replacement of the initial curvatures by the deformed ones) be
invariant under the supertransformations. As a result, we obtain the
supersymmetric vertex as a linear combination of the four elementary
ones:
\begin{equation}
\mathcal{L}_1 = g_0 V(\Omega_1,\Omega_2,\Omega_3) + 
g_1 V(\Omega_1,\Phi_2,\Phi_3) + g_2 V(\Phi_1,\Omega_2,\Phi_3) + g_3
V(\Phi_1,\Phi_2,\Omega_3).
\end{equation}
This fact drastically simplifies the construction of the
supersymmetric vertex, since most of the work has already been done in
\cite{KhZ20a}. We consider now the cases 2a, 2b and 2c in turn.

\noindent
{\bf Case 2a} corresponds to three half-integer superspin
supermultiplets $(s_i,s_i-\iz)$. Recall that all the curvatures except
the highest ones vanish on-shell; so the cubic vertex is completely
determined by the deformations of these highest curvatures. So in this
section we consider their deformations only. The deformations for the
highest curvatures for the bosonic and fermionic components of the
first supermultiplet are:
\begin{eqnarray}
\label{curv_2a}
\Delta {\mathcal{R}_1}^{\alpha(2s_1-2)} &=&
 \sum_{k=0}^{\hat{s}_1} \frac{a_0\lambda^{2k}}{k!(\hat{s}_1-k)!} 
{\Omega_2}^{\alpha(\hat{s}_3)\beta(\hat{s}_1-k)\dot\beta(k)}
{\Omega_3}^{\alpha(\hat{s}_2)}{}_{\beta(\hat{s}_1-k)\dot\beta(k)}
 \nonumber \\ 
&& + \sum_{k=0}^{\hat{s}_1-1} 
\frac{ia_1\lambda^{2k}}{k!(\hat{s}_1-k-1)!}
{\Phi_2}^{\alpha(\hat{s}_3)\beta(\hat{s}_1-k-1)\dot\beta(k)}
{\Phi_3}^{\alpha(\hat{s}_2)}{}_{\beta(\hat{s}_1-k-1)\dot\beta(k)}
 \nonumber \\ 
\Delta{\mathcal{F}_1}^{\alpha(2s_1-3)} &=&
 \sum_{k=0}^{\hat{s}_1} \frac{a_2\lambda^{2k}}{k!(\hat{s}_1-k)!} 
{\Omega_2}^{\alpha(\hat{s}_3)\beta(\hat{s}_1-k)\dot\beta(k)}
{\Phi_3}^{\alpha(\hat{s}_2-1)}{}_{\beta(\hat{s}_1-k)\dot\beta(k)}
 \\ 
&& + \sum_{k=0}^{\hat{s}_1} \frac{a_3\lambda^{2k}}{k!(\hat{s}_1-k)!} 
{\Phi_2}^{\alpha(\hat{s}_3-1)\beta(\hat{s}_1-k)\dot\beta(k)}
{\Omega_3}^{\alpha(\hat{s}_2)}{}_{\beta(\hat{s}_1-k)\dot\beta(k)}
\nonumber
\end{eqnarray}
Here and in what follows the parameters $\hat{s}_i$ are always
determined by the spins of bosons:
\begin{equation}
\hat{s}_1 = s_2 + s_3 - s_1 -1, \qquad
\hat{s}_2 = s_1 + s_3 - s_2 -1, \qquad
\hat{s}_3 = s_1 + s_2 - s_3 -1.
\end{equation}
Now we have to adjust the parameters $a_{0,1,2,3}$ so that the
deformed curvatures have correct supertransformations. Consider,
for example, the $\zeta^\alpha$ transformations for 
$\Delta {\mathcal R}_1^{\alpha(2s_1-2)}$. On the one hand, direct
calculations give us:
\begin{eqnarray}
\delta \Delta R^{\alpha(2s_1-2)} &=& 
 \sum_{k=0}^{\hat{s}_1}
\frac{iC_2\lambda^{2k+1} a_0}{k!(\hat{s}_1-k)!} 
{\Phi_2}^{\alpha(\hat{s}_3-1)\beta(\hat{s}_1-k)\dot\beta(k)}
{\Omega_3}^{\alpha(\hat{s}_2)}{}_{\beta(\hat{s}_1-k)\dot\beta(k)}\zeta^\alpha
 \nonumber \\ 
&& + \sum_{k=0}^{\hat{s}_1-1}
\frac{i\big[C_2\lambda a_0-\tilde{C}_3a_1\big]\lambda^{2k}}
{k!(\hat{s}_1-k-1)!} 
{\Phi_2}^{\alpha(\hat{s}_3)\beta(\hat{s}_1-k-1)\dot\beta(k)}
{\Omega_3}^{\alpha(\hat{s}_2)}{}_{\beta(\hat{s}_1-k)\dot\beta(k)}\zeta^\beta
 \nonumber \\ 
&& + \sum_{k=0}^{\hat{s}_1}
\frac{iC_3\lambda^{2k+1} a_0}{k!(\hat{s}_1-k)!} 
{\Omega_2}^{\alpha(\hat{s}_3)\beta(\hat{s}_1-k)\dot\beta(k)}
{\Phi_3}^{\alpha(\hat{s}_2-1)}{}_{\beta(\hat{s}_1-k)\dot\beta(k)}
\zeta^\alpha
 \nonumber \\ 
&& + \sum_{k=0}^{\hat{s}_1-1}
\frac{i\big[C_3\lambda a_0-\tilde{C}_2a_1\big]\lambda^{2k}}
{k!(\hat{s}_1-k-1)!} 
{\Omega_2}^{\alpha(\hat{s}_3)\beta(\hat{s}_1-k)\dot\beta(k)}
{\Phi_3}^{\alpha(\hat{s}_2)}{}_{\beta(\hat{s}_1-k-1)\dot\beta(k)}\zeta_\beta
\end{eqnarray}
On the other hand, we must have:
\begin{eqnarray}
\delta \Delta R^{\alpha(2s_1-2)} &=& iC_1 \Delta
{\mathcal F}_1^{\alpha(2s_1-3)} \zeta^\alpha \nonumber \\
 &=& \sum_{k=0}^{\hat{s}_1} 
\frac{C_1\lambda^{2k+1} a_3}{k!(\hat{s}_1-k)!} 
{\Phi_2}^{\alpha(\hat{s}_3-1)\beta(\hat{s}_1-k)\dot\beta(k)}
{\Omega_3}^{\alpha(\hat{s}_2)}{}_{\beta(\hat{s}_1-k)\dot\beta(k)}\zeta^\alpha
\nonumber \\
 && + \sum_{k=0}^{\hat{s}_1} 
\frac{C_1\lambda^{2k+1} a_2}{k!(\hat{s}_1-k)!} 
{\Omega_2}^{\alpha(\hat{s}_3)\beta(\hat{s}_1-k)\dot\beta(k)}
{\Phi_3}^{\alpha(\hat{s}_2-1)}{}_{\beta(\hat{s}_1-k)\dot\beta(k)}\zeta^\alpha
\end{eqnarray}
One can easily see that the two expressions are equal only if:
\begin{equation}
\tilde{C}_2a_1 = C_3\lambda a_0, \qquad
C_1a_2 = C_3a_0, \qquad
C_1a_3 = C_2a_0.
\end{equation}
Using the relations between the deformation parameters $a_{0,1,2,3}$
and the coupling constants $g_{0,1,2,3}$ (see Appendix B):
\begin{eqnarray*}
g_0 &=& 
\frac{(-1)^{s_1}(2s_1-2)!}{(\hat{s}_1)!(\hat{s}_2)!(\hat{s}_3)!}
\frac{a_0}{\lambda^{2s_1-2s_3}} \\
g_1 &=& 
\frac{(-1)^{s_1}(2s_1-2)!}{(\hat{s}_1-1)!(\hat{s}_2)!(\hat{s}_3)!}
\frac{a_1}{\lambda^{2s_1-2s_3+1}} \\
g_2 &=& 
\frac{(-1)^{s_1}(2s_1-3)!}{(\hat{s}_1)!(\hat{s}_2-1)!(\hat{s}_3)!}
\frac{a_2}{\lambda^{2s_1-2s_3}} \\
g_3 &=& -
\frac{(-1)^{s_1}(2s_1-3)!}{(\hat{s}_1)!(\hat{s}_2)!(\hat{s}_3-1)!}
\frac{a_3}{\lambda^{2s_1-2s_3-1}} 
\end{eqnarray*}
we obtain finally
\begin{equation}
\label{g_2a}
g_1 = \hat{s}_1C_2C_3g_0, \qquad
g_2 = \hat{s}_2C_1C_3g_0, \qquad
g_3 = - \hat{s}_3C_1C_2 \lambda g_0.
\end{equation}
Thus the vertex $V_3$ vanishes in the flat limit $\lambda \to 0$ in 
complete agreement with the Metsaev's results. The consideration of
generic curvature of any of the three supermultiplets is similar and
yields the same relations.

\noindent
{\bf Case 2b} is the case of two integer superspins and one
half-integer superspin, with the lowest superspin being integer. Here
we consider the case of highest-half-integer superspin, i.e. 
$(s_1,s_1-\iz)$, $(s_2,s_2+\iz)$ and $(s_3,s_3+\iz)$; the case of
highest-integer superspin is similar. In this case, the highest
curvatures deformations of the first supermultiplet have the following
form:
\begin{eqnarray}
\label{strans_2b}
\Delta {\mathcal R}_1^{\alpha(2s_1-2)} 
 &=& \sum_{k=0}^{\hat{s}_1}
\frac{a_0\lambda^{2k}}{k!(\hat{s}_1-k)!} 
{\Omega_2}^{\alpha(\hat{s}_3)\beta(\hat{s}_1-k)\dot\beta(k)}
{\Omega_3}^{\alpha(\hat{s}_2)}{}_{\beta(\hat{s}_1-k)\dot\beta(k)}
 \nonumber \\ 
&& + \sum_{k=0}^{\hat{s}_1+1}
\frac{ia_1\lambda^{2k}}{k!(\hat{s}_1-k+1)!} 
{\Psi_2}^{\alpha(\hat{s}_3)\beta(\hat{s}_1-k+1)\dot\beta(k)}
{\Psi_3}^{\alpha(\hat{s}_2)}{}_{\beta(\hat{s}_1-k+1)\dot\beta(k)}
 \nonumber \\ 
\Delta {\mathcal F}_1^{\alpha(2s_1-3)} 
 &=& \sum_{k=0}^{\hat{s}_1+1} 
\frac{a_2\lambda^{2k}}{k!(\hat{s}_1-k+1)!}  
{\Omega_2}^{\alpha(\hat{s}_3-1)\beta(\hat{s}_1+1-k)\dot\beta(k)}{\Psi_3}^{\alpha(\hat{s}_2)}{}_{\beta(\hat{s}_1+1-k)\dot\beta(k)}
  \\ 
&& + \sum_{k=0}^{\hat{s}_1+1} 
\frac{a_3\lambda^{2k}}{k!(\hat{s}_1-k+1)!} 
{\Psi_2}^{\alpha(\hat{s}_3)\beta(\hat{s}_1-k+1)\dot\beta(k)}
{\Omega_3}^{\alpha(\hat{s}_2-1)}{}_{\beta(\hat{s}_1-k+1)\dot\beta(k)}
\nonumber
\end{eqnarray}
Again, the supercovariance conditions fix $a_i$, $i=0,1,2,3$ up to the
total factor; their expressions are slightly different from the case
2a:
\begin{equation}
\tilde{C}_2\lambda a_1 = - C_3a_0, \qquad
\tilde{C}_2a_1 = - C_1a_2, \qquad
\tilde{C}_3a_1 = C_1a_3.
\end{equation}
This leads to the following relations for the coupling constants:
\begin{equation}
(\hat{s}_1+1)g_1 = - C_2C_3g_0, \qquad
(\hat{s}_1+1)g_2 = \hat{s}_3C_1C_3 \lambda g_0, \qquad
(\hat{s}_1+1)g_3 = \hat{s}_2C_1C_2g_0.
\end{equation}
This time the vertex $V_2$ vanishes in the flat limit again in
agreement with the Metsaev's results.\\
\noindent
{\bf Case 2c} is the case of two integer superspins and one
half-integer superspin, with the lowest spin being half-integer, i.e.
$(s_1,s_1+\iz)$, $(s_2,s_2+\iz)$ and $(s_3,s_3-\iz)$. 
The expressions for the highest curvatures deformations of the first
supermultiplets now have the following form:
\begin{eqnarray}
\Delta {\mathcal R}_1^{\alpha(2s_1-2)} &=&
 \sum_{k=0}^{\hat{s}_1} 
\frac{a_0\lambda^{2k}}{k!(\hat{s}_1-k)!} 
{\Omega_2}^{\alpha(\hat{s}_3)\beta(\hat{s}_1-k)\dot\beta(k)}
{\Omega_3}^{\alpha(\hat{s}_2)}{}_{\beta(\hat{s}_1-k)\dot\beta(k)}
 \nonumber \\ 
&& + \sum_{k=0}^{\hat{s}_1} 
\frac{ia_1\lambda^{2k}}{k!(\hat{s}_1-k)!}  
{\Psi_2}^{\alpha(\hat{s}_3+1)\beta(\hat{s}_1-k)\dot\beta(k)}
{\Phi_3}^{\alpha(\hat{s}_2-1)}{}_{\beta(\hat{s}_1-k)\dot\beta(k)}
 \nonumber \\ 
\Delta {\mathcal F}_1^{\alpha(2s_1-1)} &=&
 \sum_{k=0}^{\hat{s}_1-1} 
\frac{a_2\lambda^{2k}}{k!(\hat{s}_1-k-1)!} 
{\Omega_2}^{\alpha(\hat{s}_3+1)\beta(\hat{s}_1-k-1)\dot\beta(k)}{\Phi_3}^{\alpha(\hat{s}_2)}{}_{\beta(\hat{s}_1-k-1)\dot\beta(k)}
  \\ 
&& + \sum_{k=0}^{\hat{s}_1} 
\frac{a_3\lambda^{2k}}{k!(\hat{s}_1-k)!} 
{\Psi_2}^{\alpha(\hat{s}_3+1)\beta(\hat{s}_1-k)\dot\beta(k)}
{\Omega_3}^{\alpha(\hat{s}_2)}{}_{\beta(\hat{s}_1-k)\dot\beta(k)}
\nonumber
\end{eqnarray}
In this case the supercovariance conditions are:
\begin{eqnarray}
\delta {\mathcal F}_1^{\alpha(2s_1-1)} &=&
\tilde{C}_1  {\mathcal{R}_1}^{\alpha(2s_1-2)} \zeta^{\alpha}
 \nonumber \\ 
\delta{\mathcal R}_1^{\alpha(2s_1-2)} &=&
iC_1 {\mathcal{F}_1}^{\alpha(2s_1-2)\dot\alpha} \zeta_{\dot\alpha}
 + iC_1 {\mathcal{F}_1}^{\alpha(2s_1-2)\beta} \zeta_\beta.
\end{eqnarray}
Again, the factors $a_i$, $i=0,1,2,3$ are fixed by the supercovariance
conditions up to the total factor:
\begin{equation}
C_2a_0 = C_1a_3, \qquad
\tilde{C}_1a_1 = C_3a_3, \qquad
C_2a_2 = C_3\lambda a_3,
\end{equation}
while for the coupling constants we obtain now:
\begin{equation}
g_0 = (\hat{s}_3+1)\tilde{C}_1\tilde{C}_2 \lambda g_3, \qquad 
g_1 = \hat{s}_2\tilde{C}_1C_3g_3, \qquad
g_2 = - \hat{s}_1\tilde{C}_2C_3g_3,
\end{equation}
thus the bosonic vertex $V_0$ vanishes in the flat limit as it should
be.

\section{Supersymmetric vertices in the flat space}

The deformation procedure we used in the previous section to construct
cubic vertices in $AdS_4$ generates terms having up to $N_{max} =
s_1+s_2+s_3-2$ derivatives, while the corresponding flat vertex must
have $N_b = s_1+s_2-s_3$ for the bosonic case and $N_f =
s_1+s_2-s_3-1$ for the fermionic one. In space-time with $D > 4$ these
higher derivatives terms reproduce the so-called abelian vertices
\cite{Vas11}, which are absent in $d=4$. In {\cite{KhZ20a} we have
shown that all these higher derivatives terms can be combined in the
total derivatives or vanish on-shell. This allowed us to obtain a 
non-singular flat limit.

In the same way we can consider the flat limit of the supersymmetric
vertices. The supertransformations we used also admit a non-singular
flat limit, so one can expect that the flat vertex also must be
supersymmetric. However, as will be seen later on, it is important to
use the full supertransformations (\ref{strans_23_int}) and
(\ref{strans_23_halfint}), i.e. with generalized Weyl tensor terms.
This is due to the fact that we use mass shell equations to simplify
the vertex; the supertransformations have to preserve these equations,
which is impossible without Weyl tensor terms. 
In the previous section we have seen that in $AdS_4$ any
such vertex consists of the four elementary ones, but one of them
vanishes in the flat limit. As we have shown in \cite{KhZ20a}, the
flat vertices have the most simple form when all three spins are
different. So in  this section we provide as an illustration the
examples of the supersymmetric vertices restricting ourselves with
their simplest representatives. 

Now we consider the cases 2a, 2b and 2c in turn. \\
{\bf Case 2a} In general such vertex has the following form:
\begin{equation}
V = ig_0 V_0(\Omega_1,\Omega_2,\Omega_3)
+ g_1 V_1(\Omega_1,\Phi_2,\Phi_3)
 + g_2 V_2(\Phi_1,\Omega_2,\Phi_3).
\end{equation}
In case of sufficiently different spins, i.e. if $s_1>s_2>s_3$, the
vertex can be written as:
\begin{eqnarray}
V_{2a} &=& ig_0 d\Omega_{3,\alpha(\hat{s}_2)\beta(\hat{s}_1)}
\Omega_1^{\alpha(\hat{s}_2)\dot\alpha(\hat{s}_3)}
\Omega_2^{\beta(\hat{s}_1)}{}_{\dot\alpha(\hat{s}_3)}
 \nonumber \\
 && + g_1 d \Phi_{3,\alpha(\hat{s}_2)\beta(\hat{s}_1-1)}
\Omega_1^{\alpha(\hat{s}_2)\dot\alpha(\hat{s}_3)}
\Phi_2^{\beta(\hat{s}_1-1)}{}_{\dot\alpha(\hat{s}_3)}
 \nonumber \\ 
&& + g_2 d \Phi_{3,\alpha(\hat{s}_2-1)\beta(\hat{s}_1)}
\Phi_1^{\alpha(\hat{s}_2-1)\dot\alpha(\hat{s}_3)}
\Omega_2^{\beta(\hat{s}_1)}{}_{\dot\alpha(\hat{s}_3)}
 + h.c.
\end{eqnarray}
We pick this case to demonstrate how the invariance under the
supertransformations is achieved. Let us begin with 
$\zeta^\alpha$-transformations:
\begin{eqnarray*}
\delta \Omega_1^{\alpha(\hat{s}_2)\dot\alpha(\hat{s}_3)} &=& 
iC_1 \Phi_1^{\alpha(\hat{s}_2-1)\dot\alpha(\hat{s}_3)} \zeta^\alpha \\
\delta \Omega_2^{\beta(\hat{s}_1)}{}_{\dot\alpha(\hat{s}_3)} &=& 
iC_2 \Phi_2^{\beta(\hat{s}_1-1)}{}_{\dot\alpha(\hat{s}_3)} \zeta^\beta
\\
\delta \Phi_{3,\alpha(\hat{s}_2)\beta(\hat{s}_1-1)} &=& - \tilde{C}_3
\Omega_{3,\alpha(\hat{s}_2)\beta(\hat{s}_1-1)\gamma} \zeta^\gamma \\
\delta \Phi_{3,\alpha(\hat{s}_2-1)\beta(\hat{s}_1)} &=& - \tilde{C}_3
\Omega_{3,\alpha(\hat{s}_2-1)\beta(\hat{s}_1)\gamma} \zeta^\gamma
\end{eqnarray*}
Calculating the variation of the cubic vertex we obtain:
\begin{eqnarray*}
\delta V_{2a} &=& [\tilde{C}_3g_2 - \hat{s}_2C_1g_0]
d \Omega_{3,\alpha(\hat{s}_2-1)\beta(\hat{s}_1)\gamma}
\Phi_1^{\alpha(\hat{s}-2-1)\dot\alpha(\hat{s}_3)}
\Omega_2^{\beta(\hat{s}_1)}{}_{\dot\alpha(\hat{s}_3)} \zeta^\gamma
\\
 && + [\tilde{C}_3g_1 - \hat{s}_1C_2g_0]
d \Omega_{3,\alpha(\hat{s}_2)\beta(\hat{s}_1-1)\gamma}
\Omega_1^{\alpha(\hat{s}_2)\dot\alpha(\hat{s}_3)}
\Phi_2^{\beta(\hat{s}_2-1)}{}_{\dot\alpha(\hat{s}_3)} \zeta^\gamma
\\
 && + i[\hat{s}_1C_2g_2 - \hat{s}_2C_1g_1]
d \Phi_{3,\alpha(\hat{s}_2-1)\beta(\hat{s}_1-1)\gamma}
\Phi_1^{\alpha(\hat{s}_2-1)\dot\alpha(\hat{s}_3)}
\Phi_2^{\beta(\hat{s}_1-1)}{}_{\dot\alpha(\hat{s}_3)} \zeta^\gamma
\end{eqnarray*}
This variation vanishes due to the relations (\ref{g_2a}). Now
consider $\zeta^{\dot\alpha}$-transformations:
\begin{eqnarray*}
\delta \Omega_{3,\alpha(\hat{s}_2)\beta(\hat{s}_1)} &=& 
iC_3 e^{\gamma\dot\alpha} Y_{\alpha(\hat{s}_2)\beta(\hat{s}_1)\gamma}
\zeta_{\dot\alpha} 
\\
\delta \Phi_1^{\alpha(\hat{s}_1-1)\dot\alpha(\hat{s}_3)} &=& 
\tilde{C}_1 
\Omega_1^{\alpha(\hat{s}_1-1)\dot\alpha(\hat{s}_3)\dot\beta}
\zeta_{\dot\beta} 
\\
\delta \Phi_2^{\beta(\hat{s}_1-1)}{}_{\dot\alpha(\hat{s}_3)} &=&
- \tilde{C}_2
\Omega_2^{\beta(\hat{s}_1-1)}{}_{\dot\alpha(\hat{s}_3)\dot\beta}
\zeta^{\dot\beta}
\end{eqnarray*}
This produces:
\begin{eqnarray*}
\delta {\cal L}_1 &=& - C_3g_0 e^{\gamma\dot\beta}
dY_{\alpha(\hat{s}_2)\beta(\hat{s}_1)\gamma} \zeta_{\dot\beta}
\Omega_1^{\alpha(\hat{s}_2)\dot\alpha(\hat{s}_3)}
\Omega_2^{\beta(\hat{s}_1)}{}_{\dot\alpha(\hat{s}_3)} 
\\
 && + \tilde{C}_1g_2
d \Phi_{3,\alpha(\hat{s}_2-1)\beta(\hat{s}_1)}
\Omega_1^{\alpha(\hat{s}_2-1)\dot\alpha(\hat{s}_3)\dot\beta} 
\zeta_{\dot\beta}
\Omega_2^{\beta(\hat{s}_1)}{}_{\dot\alpha(\hat{s}_3)} 
\\
 && - \tilde{C}_2g_1
d \Phi_{3,\alpha(\hat{s}_2)\beta(\hat{s}_1-1)}
\Omega_1^{\alpha(\hat{s}_2)\dot\alpha(\hat{s}_3)}
\Omega_2^{\beta(\hat{s}_1-1)}{}_{\dot\alpha(\hat{s}_3)\dot\beta}
\zeta^{\dot\beta}
\end{eqnarray*}
Using the on-shell identities, in-particular
$$
d \Phi_{3,\alpha(2s_3-3)} \approx -
E^{\beta(2)} Y_{\alpha(2s_3-3)\beta(2)}
$$
and the fact that the Lagrangian is defined up to total derivative,
one can transform $\delta V_{2a}$ to the following expression:
\begin{eqnarray*}
\delta V_{2a} &=& [ \hat{s}_2C_3g_0 - \tilde{C}_1g_2] E^{\gamma(2)}
Y_{\alpha(\hat{s}_2-1)\beta(\hat{s}_1)\gamma(2)}
\Omega_1^{\alpha(\hat{s}_2-1)\dot\alpha(\hat{s}_3)\dot\beta}
\Omega_2^{\beta(\hat{s}_1)}{}_{\dot\alpha(\hat{s}_3)}
\zeta_{\dot\beta} 
\\
 && + [ \tilde{C}_2g_1 - \hat{s}_1C_3g_0]E^{\gamma(2)}
Y_{\alpha(\hat{s}_2)\beta(\hat{s}_1-1)\gamma(2)}
\Omega_1^{\alpha(\hat{s}_2)\dot\alpha(\hat{s}_3)}
\Omega_2^{\beta(\hat{s}_-1-1)}{}_{\dot\alpha(\hat{s}_3\dot\beta}
\zeta^{\dot\beta} 
\end{eqnarray*}
This expression also vanishes  due to the relations (\ref{g_2a}).

\noindent
{\bf Case 2b}  We consider the case of superspins
$Y_{s_1-\iz},Y_{s_2},Y_{s_3}$ with $s_2>s_3$. The general form of the
flat vertex is:
\begin{equation}
V_{2b} = ig_0 V_0(\Omega_1,\Omega_2,\Omega_3)
+ g_1 V_1(\Omega_1,\Psi_2,\Psi_3)
+ g_3 V_3(\Phi_1,\Psi_2,\Omega_3)
\end{equation}
In case of sufficiently different spins it can be written as:
\begin{eqnarray}
V_{2b} &=& ig_0 D \Omega_{3,\alpha(\hat{s}_2)\beta(\hat{s}_1)}
\Omega_1^{\alpha(\hat{s}_2)\dot\alpha(\hat{s}_3)}
\Omega_2^{\beta(\hat{s}_1)}{}_{\dot\alpha(\hat{s}_3)}
 \nonumber \\ 
&& + g_1 D \Psi_{3,\alpha(\hat{s}_2)\beta(\hat{s}_1+1)}
\Omega_1^{\alpha(\hat{s}_2)\dot\alpha(\hat{s}_3)}
\Psi_2^{\beta(\hat{s}_1+1)}{}_{\dot\alpha(\hat{s}_3)}
 \nonumber \\ 
&& + g_3 D \Omega_{3,\alpha(\hat{s}_2-1)\beta(\hat{s}_1+1)}
\Phi_1^{\alpha(\hat{s}_2-1)\dot\alpha(\hat{s}_3)}
\Psi_2^{\beta(\hat{s}_1+1)}{}_{\dot\alpha(\hat{s}_3)} + h.c.
\end{eqnarray}

\noindent
{\bf Case 2c}  This case corresponds to superspins 
$Y_{s_1},Y_{s_2},Y_{s_3-\iz}$. The general form of the flat vertex is:
\begin{equation}
V_{2c} = g_1 V_1(\Omega_1,\Psi_2,\Phi_3)
+ g_2 V_2(\Psi_1,\Omega_2,\Phi_3)
+ g_3 V_3(\Psi_1,\Psi_2,\Omega_3)
\end{equation}
In case of sufficiently different spins, the vertex has the form:
\begin{eqnarray}
V_{2c} &=& g_1 D \Phi_{3,\alpha(\hat{s}_2-1)\beta(\hat{s}_1)}
\Omega_1^{\alpha(\hat{s}_2-1)\dot\alpha(\hat{s}_3+1)}
\Psi_2^{\beta(\hat{s}_1)}{}_{\dot\alpha(\hat{s}_3+1)}
 \nonumber \\ 
&& + g_2 D \Phi_{3,\alpha(\hat{s}_2)\beta(\hat{s}_1-1)}
\Psi_1^{\alpha(\hat{s}_2)\dot\alpha(\hat{s}_3+1)}
\Omega_2^{\beta(\hat{s}_1-1)}{}_{\dot\alpha(\hat{s}_3+1)}
 \nonumber \\ 
&& + g_3 D \Omega_{3,\alpha(\hat{s}_2)\beta(\hat{s}_1)}
\Psi_1^{\alpha(\hat{s}_2)\dot\alpha(\hat{s}_3+1)}
\Psi_2^{\beta(\hat{s}_1)}{}_{\dot\alpha(\hat{s}_3+1)} + h.c.
\end{eqnarray}

\section{Conclusion}

We have constructed supersymmetric cubic vertices in the 
four-dimensional space for three massless supermultiplets with all
their spins obeying the strict triangle inequality, $s_1<s_2+s_3$. 
Our procedure is a straightforward generalisation of the so-called
Fradkin-Vasiliev formalism to the case of massless higher spin
supermultiplets and it also based on our previous results in
\cite{KhZ20a}. First of all, we construct supersymmetric vertices in
$AdS$ space and show that each such vertex necessarily contains all
four (what we call) elementary vertices. At the same time, in the flat
limit $\lambda \to 0$ one of the elementary vertices always vanishes
in complete agreement with the Metsaev's classification \cite{Met19a}.
Wt provide simple examples of the flat supersymmetric vertices and
directly check that they are invariant under the supertransformations.

\section*{Acknowledgements}
M.Kh. is grateful to Foundation for the Advancement of Theoretical
Physics and Mathematics "BASIS" for their support of the work.

\appendix

\section{Notation and conventions}

We use the same notations and conventions as in our previous works
\cite{KhZ20a,KhZ20}. First of all, all objects are the multispinors in
their local indices $\Phi^{\alpha(k)\dot\alpha(l)}$, where
$\alpha,\dot\alpha = 1,2$ and $k,l$ denote the number of completely
symmetric undotted and dotted indices, respectively. All indices
denoted by the same letter and placed on the same level are assumed to
be symmetrized, where symmetrization is understood as the sum of the
minimal number of terms necessary without normalization factor.

In the frame-like formalism, which we use, the fields are one-forms.
We omit the wedge product signs for brevity. We work
in $AdS_4$ space, characterized by a cosmological constant
$\Lambda=-\lambda^2$ and its flat limit, described by the background
frame $e^{\alpha\dot{\alpha}}$, which is one-form, and by the
(exterior) background Lorentz covariant derivative $D$, for which the
following identities hold:
$$
De^{\alpha\dot{\alpha}} = 0,\qquad 
D^2 \Phi^{\alpha(k)\dot\alpha(l)} = - \lambda^2
\big[ E^\alpha{}_\beta \Phi^{\alpha(k-1)\beta\dot\alpha(l)} +
E^{\dot{\alpha}}{}_{\dot\beta} 
\Phi^{\alpha(k)\dot\alpha(l-1)\dot\beta}\big]
$$
Here $E^{\alpha(2)}$ and $E^{\dot\alpha(2)}$ are the components of
the background two-forms defined by the following identity:
$$
e^{\alpha\dot\alpha} e^{\beta\dot\beta} =
\epsilon^{\alpha\beta} E^{\dot\alpha\dot\beta} +
\epsilon^{\dot\alpha\dot\beta} E^{\alpha\beta}
$$

\section{Elementary cubic vertices}

For concreteness we consider here the case of the bosonic vertices. As
we have already mentioned, the first step in the construction is find
such quadratic deformation of all gauge invariant curvatures
$\Delta {\cal R} \sim \Omega\Omega$ that the deformed curvatures
$\hat{\cal R} = {\cal R} + \Delta {\cal R}$ transform covariantly
under the (deformed) gauge transformations:
$\delta \hat{\cal R} \sim {\cal R} \eta$. This step is carried out for
each three fields independently, so let us give here the result for
the first one:
\begin{eqnarray}
\Delta {\mathcal R}_1^{\alpha(2s_1-2-m)\dot\alpha(m)} &=&
\sum_{k=0}^{\hat{s}_1}\sum_{l=0}^{min(m,\hat{s}_2)}
\frac{a}{(\hat{s}_1-k)!k!} 
\nonumber \\
&& \times 
{\Omega_2}^{\alpha(\hat{s}_3-m+l)\beta(\hat{s}_1-k)\dot\alpha(m-l)\dot{\beta}(k)}
{\Omega_3}^{\alpha(\hat{s}_2-l)\dot\alpha(l)}{}_{\beta(\hat{s}_1-k)\dot\beta(k)}
\end{eqnarray}
Here we introduced convenient combinations
$$
\hat{s}_1 = s_2 + s_3 - s_1 - 1, \qquad
\hat{s}_2 = s_1 + s_3 - s_2 - 1, \qquad
\hat{s}_3 = s_1 + s_2 - s_3 - 1.
$$
Note that for spins $s_{1,2,3}$  satisfying the strict triangular
inequality these combinations are non-negative. Note also that they
are always integer even then two of the particles are fermions. As we
see, the deformation is determined up to one arbitrary parameter $a$;
we denote $b$ and $c$ the corresponding parameters for the second and
third particles.

The second step is to take the sum of the three Lagrangians  with the
initial curvatures replaced by the deformed ones and require it to be
gauge invariant on-shell. This leads to the following relations:
\begin{equation}
\frac{(-1)^{s_1} (2s_1-2)!}{\lambda^{2s_1-2}}a 
= \frac{(-1)^{s_1+s_3} (2s_2-2)!}{\lambda^{2s_2-2}}b
 = \frac{(-1)^{s_3} (2s_3-2)!}{\lambda^{2s_3-2}}c
\end{equation}

Such procedure produces cubic terms with the number of derivatives up
to $N_{max} = s_1+s_2+s_3-2$, while the corresponding flat space
analogue has only $N_0 = s_1+s_2-s_3$. In space-time with $d > 4$
these higher derivatives terms would reproduce the so-called abelian
vertices \cite{Vas11}, which are absent in $d=4$. As we have shown in
\cite{KhZ20a}, all the terms with $N > N_0$ derivatives can be
combined into total derivative or vanish on-shell. This allows us to
get a non-singular flat limit $\lambda \to 0$ and obtain the
corresponding flat versions. If $g$ is the coupling constant for such
vertex, then for the deformation parameters $a,b,c$ we have, for
example:
\begin{equation}
\label{g_def}
\frac{(-1)^{s_1} (2s_1-2)!}{(\hat{s}_1)!(\hat{s}_2)!(\hat{s}_3)!} a =
\lambda^{2s_1-2s_3}  g
\end{equation}


\begin{thebibliography}{10}

\bibitem{Met19a}
R.R. Metsaev
{\it "Cubic interaction vertices for N=1 arbitrary spin massless
  supermultiplets in flat space",}
JHEP {\bf 08} (2019) 130, arXiv:1905.11357.

\bibitem{Met19b}
R.R. Metsaev
{\it "Cubic interactions for arbitrary spin N-extended massless
supermultiplets in 4d flat space",}
JHEP {\bf 11} (2019) 084, arXiv:1909.05241.

\bibitem{Met05}
R.~R. Metsaev
{\it "Cubic interaction vertices of massive and massless higher spin
fields",}
Nucl. Phys. {\bf B759} (2006) 147, arXiv:hep-th/0512342.

\bibitem{Met07b}
R.~R. Metsaev
{\it "Cubic interaction vertices for fermionic and bosonic arbitrary
spin fields",}
Nucl. Phys. {\bf B859} (2012) 13, arXiv:0712.3526.

\bibitem{BGK17}
I.~L. Buchbinder, S.~James~Gates Jr., Konstantinos Koutrolikos
{\it "Higher Spin Superfield interactions with the Chiral
Supermultiplet: Conserved Supercurrents and Cubic Vertices",}
Universe {\bf 4} (2018) 6, arXiv:1708.06262.

\bibitem{BGK18b}
I.~L. Buchbinder, S.~James~Gates Jr., K.~Koutrolikos
{\it "Integer superspin supercurrents of matter supermultiplets",}
JHEP {\bf 05} (2019) 031, arXiv:1811.12858.

\bibitem{BGK18a}
I.~L. Buchbinder, S.~James~Gates Jr., Konstantinos Koutrolikos
{\it "Conserved higher spin supercurrents for arbitrary spin massless
  supermultiplets and higher spin superfield cubic interactions",}
JHEP {\bf 08} (2018) 055, arXiv:1805.04413.

\bibitem{GK19}
S.~James~Gates Jr., K.~Koutrolikos
{\it "Progress on cubic interactions of arbitrary superspin
supermultiplets via gauge invariant supercurrents",}
Phys. Lett. {\bf B797} (2019) 134868, arXiv:1904.13336.

\bibitem{KhZ20a}
M.~V. Khabarov, Yu.~M. Zinoviev
{\it "Massless higher spin cubic vertices in flat four dimmensional
space",}
JHEP {\bf 08} (2020) 112, arXiv:2005.09851.

\bibitem{FV87}
E.~S. Fradkin, M.~A. Vasiliev
{\it "On the gravitational interaction of massless higher-spin
fields",}
Phys. Lett. {\bf B189} (1987) 89.

\bibitem{FV87a}
E.~S. Fradkin, M.~A. Vasiliev
{\it "Cubic interaction in extended theories of massless higher-spin
fields",}
Nucl. Phys. {\bf B291} (1987) 141.

\bibitem{Vas11}
M.~Vasiliev
{\it "Cubic Vertices for Symmetric Higher-Spin Gauge Fields in
$(A)dS_d$",}
Nucl. Phys. {\bf B862} (2012) 341, arXiv:1108.5921.

\bibitem{KhZ20}
M.~V. Khabarov, Yu.~M. Zinoviev
{\it "Massive higher spin supermultiplets unfolded",}
Nucl. Phys. {\bf B953} (2020) 114959, arXiv:2001.07903.

\end{thebibliography}
\end{document}